\documentclass[10pt]{iopart}

\usepackage{graphicx}
\usepackage{subfigure}
\usepackage{iopams} 

\begin{document}

\title[A simple model for the growth of animal cells in culture]{Cluster-cluster aggregation with particle replication and chemotaxy: a simple model for the growth of animal cells in culture}

\author{S. G. Alves$^1$, M. L. Martins$^{1,2}$}
\address{$^1$Departamento de F\'{\i}sica, Universidade Federal de Vi\c{c}osa, \\
36570-000, Vi\c{c}osa, MG, Brazil}
\address{$^2$National Institute of Science and Technology for Complex Systems}
\eads{sidiney@ufv.br}

\begin{abstract}
Aggregation of animal cells in culture comprises a series of motility, collision and adhesion processes of basic relevance for tissue engineering, bioseparations, oncology research and \textit{in vitro} drug testing. In the present paper, a cluster-cluster aggregation model with stochastic particle replication and chemotactically driven motility is investigated as a model for the growth of animal cells in culture. The focus is on the scaling laws governing the aggregation kinetics. Our simulations reveal that in the absence of chemotaxy the mean cluster size and the total number of clusters scale in time as stretched exponentials dependent on the particle replication rate. Also, the dynamical cluster size distribution functions are represented by a scaling relation in which the scaling function involves a stretched exponential of the time. The introduction of chemoattraction among the particles leads to distribution functions decaying as power laws with exponents that decrease in time. The fractal dimensions and size distributions of the simulated clusters are qualitatively discussed in terms of those determined experimentally for several normal and tumoral cell lines growing in culture. It is shown that particle replication and chemotaxy account for the simplest cluster size distributions of cellular aggregates observed in culture.

\end{abstract}

{\bf Accepted for publication: {Jstat}}
\maketitle

\section{Introduction}

Recent experiments \cite{Rose,Vilela} investigated the dynamics of normal and cancer cells in culture through the cluster size distribution function. These distributions for the human malignant cell lines HN-5 (carcinoma of the head and neck), MCF-10 (normal of the breast) and MDA-MB-231 (breast carcinoma) scaled as power laws at any time in culture. Consequently, HN-5, MCF-10, and MB-231 cell aggregates had no characteristic sizes. For MCF-7 cells (breast carcinoma), a transition from a power law to an exponential decay of this cluster size distribution function occurred around 96 h in culture. This transition was associated to the weakening of cell-cell adhesion. Thus, after the transition, MCF-7 aggregates exhibit a characteristic size and their larger clusters progressively break down into smaller ones. Inverse transitions, i.e., from an exponential to a power law scale, were observed in the aggregation regimes of MDCK (canine normal) and Hep-2 (human carcinoma of the larynx) cells. Furthermore, in MDCK cells the phenotypic transition is irreversible. Indeed, cell colonies generated by MDCK cells cloned after the transition exhibit cluster size distributions with a power law decay at any time in culture. In contrast, colonies cloned from cells before the transition show the same behavior as the original culture, particularly the transition in their aggregation kinetics. These transitions are cell dependent responses to face increasing inhibitory constraints to growth. For MCF-7 cells the transformation leads to increased migration through a less adhesive phenotype, whereas for MDCK and Hep-2 cells the transitions select strong adhesive phenotypes.

A main question is if such transitions indicate the expression of incrementally transformed states favored by their enhanced capacity to grow under inhibitory conditions. There is increasing evidence that cancer progression is initially driven by selection of transformed cell phenotypes that confer proliferative advantages over the competing normal populations \cite{Rubin,Chow}. In particular, Rubin \cite{Rubin} demonstrated that ``focus formation'' in NIH 3T3 cells is a neoplastic transformation in culture primarily based on the selection of clones derived from single cells. In such clones, successive mutations and rounds of further selection culminate in genetic variants able to proliferate better than the surrounding cells in a confluent culture or under other constraints that inhibit or slow the multiplication of normal cells. As one can notice, the study of cell growth in culture can provide valuable insights about neoplastic transformation, cancer progression and metastasis \cite{Chambers}, but also on normal physiological processes as, for instance, embryonic development \cite{Keller}, immune surveillance and wound healing \cite{Friedl}. Moreover, culture assays allow us to accurately measure cell aggregation kinetics (positions, velocities, aggregation rates, cluster morphology, cell-cell correlations etc.) and probe the biological mechanisms (adhesion processes, cytoskeletal activity, cell signalling and polarization etc.) that affect cellular aggregation. Kinetic parameters can distinguish differences in the motility and aggregation of distinct cell lines which may have prognostic value in oncology.

Interestingly, despite their potential, relatively little attention has been directed to the experimental characterization and theoretical modeling of cell growth in culture through the methods of kinetic aggregation and scaling analysis. Kinetic models based on Smoluchowski theory were used to determine aggregation rates and diffusion coefficients of multicellular spheroids formed by prostate cancer cells \cite{Enmon1,Enmon2}. The cluster-cluster aggregation model (CCA) \cite{Meakin,Kolb} with varying sticking probability, size-independent diffusion rates, and internal relaxations was applied to simulate restructuring DU 145 and LNCaP prostate cancer spheroids \cite{Connor}. More recently, the formation of networks of interconnected clusters in cardiac cell culture was analyzed \cite{Harada}. The spatial distribution of cellular clusters, the radial distribution functions for cell nuclei, and the cluster-size distribution functions for ventricular cells from neonatal rats were determined. The results suggest that cell aggregation is induced by the relocation of focal contacts triggered by the periodic contraction of these cells. At last, through a completely distinct strategy that does not involve cell proliferation or apoptosis, human alveolar type II epithelial cells form cysts in 3D Matrigel cultures. Again, these alveolar-like cysts were simulated via a CCA extended model in which the particles (cells) are autonomous agents interacting among them (attachment/rearrangement) and with their surrounding environment (chemotaxy) \cite{Kim}. The resulting self-organized structures closely resemble the \textit{in vitro} alveolar-like cyst morphologies and size distributions.

In the present paper we propose a first order model to describe the growth of animal cells in monolayer cultures. In this algorithmic approach the dynamics of cell migration, aggregation and division is modeled through a CCA model with particle replication and chemotactically driven motility. Our main goal is determine the scaling laws generated by this extended CCA process and compare them with those observed in culture assays. Neatly, a more realistic and quantitative model for cells in culture must include several complex biological phenomena such as cell-cell adhesion, collective cell polarization and migration that were neglected here. The paper is organized as follows. In Section \ref{modelo} the CCA model with particle replication and chemotaxy is introduced. Section \ref{resultado} shows and discuss the results obtained from numerical simulations. Finally, in Section \ref{conclusao} some conclusions are drawn.

\section{Model and methods}
\label{modelo}
Animal cells seeded in a culture plate form colonies as time evolves. These colonies emerge even from very low initial cell densities described by a distribution primarily composed of single cells \cite{Rose,Vilela}. In culture, individual cells multiply by division and begin to form small clusters. Simultaneously, cells migrate and aggregate with other cells that may have already divided. As a result of this, clusters enlarge not only by new cell divisions but also by the aggregation of new cells that move towards them and, when clusters are big enough, by the merging of adjacent clusters. The growth patterns of animal cell lines in culture can be characterized using, for instance, the fractal dimension \cite{Sandro} and the cluster size distribution of cellular aggregates \cite{Rose,Vilela}.

From the optics of a physicist, the scaling laws measured in culture assays must be contrasted with those obtained from mathematical models intended to describe this phenomenon. The aforementioned general scenario for the dynamics of cells in culture evokes the CCA model as the simplest candidate to modeling such complex growth process. Furthermore, at least cell division, the most basic biological feature not taken into account in the original CCA version, should be included in a zeroth model for cells in culture.

Our model straightforwardly extends the CCA algorithm in order to describe the aggregation of self-replicating particles driven by diffusion under chemotactic cell signaling. Initially, a low number $N_0$ of particles, representing cells of unitary diameter, is randomly distributed over a square lattice of linear size $L$. Periodic boundary conditions are used in order to reduce finite-size effects. The double occupation of any lattice site is always forbidden. A set of $s$ particles interconnected through a sequence of nearest neighbor particles forms a cluster of size $s$. Thus, single particles define $s=1$ clusters, interconnected pairs form $s=2$ clusters and so on. The initial cluster distribution changes in time according to the following evolution rules. At each time step a cluster is randomly selected with equiprobability and performs, again with equal chance, one of two actions: migration or replication.

When migration is chosen, the cluster can move rigidly by one lattice unit with probability $p_{mov}$. As in the original CCA model, this probability is given by

\begin{equation}
p_{mov} = \frac{D_s} {D_{max}}.
\end{equation}
Here $D_s = C s^\gamma$ is the diffusion coefficient of a cluster of size $s$ and $D_{max}$ is the largest diffusion coefficient. So, the mobility of a cluster depends on its size $s$, and this dependence is controlled by the exponent $\gamma$. In particular, the $\gamma = 0$ case corresponds to a mass-independent diffusion coefficient, whereas $\gamma=-1/d$ ($d$ being the spatial dimension) corresponds to a diffusion coefficient inversely proportional to the cluster mass or gyration radius (Stokes' law).

In turn, when replication is the chosen action, all peripheral particles of the cluster can self-replicate (or not) with a probability $p_{div}$ ($1-p_{div}$). At random, each peripheral particle is sequentially selected and can self-replicate if at least one of its nearest neighbor sites is empty. Its replica will occupy randomly one of such empty sites, remaining permanently stuck to the cluster. After each tentative action (migration or replication) the time is incremented by $\Delta t=1/N(t)$, where $N(t)$ is the total number of clusters present in the system at time $t$. Also, if the resulting cluster (at a new location or containing additional particles) becomes adjacent to another one, they stick together forming a larger cluster. Adjacent clusters are connected through at least one pair of nearest neighbor particles, one of them from the first cluster and the other from the second cluster.

In order to model the chemotactically driven migration of cells in culture, it was assumed that each cell synthesizes a diffusive chemoattractant signal whose concentration field $C$ obeys the diffusion equation
\begin{equation}
\frac{dC}{dt}= D \nabla^2 C + \lambda \Gamma_{i,j} - \nu C\nonumber
\end{equation}
where $D$ is the signal diffusion coefficient, $\lambda$ its production rate and $\nu$ a parameter controlling its natural degradation.

Concerning the analysis of this generalized CCA model involving two major extensions, particle replication and chemotaxy, it is more convenient to consider the effects of such extensions incrementally. So, three different variants of the model were separately investigated, namely:
\begin{itemize}
\item Model I - CCA with particle replication. In this version cluster-cluster aggregation with particle replication under a purely diffusion-limited regime is considered.
\item Model II - Chemotactically driven CCA. The movement of the clusters are guided by the cell signaling, but there is no particle replication.
\item Model III - CCA with particle replication and chemotaxy. Here, both particle replication and cluster motility driven by cell signaling (chemotaxis) are considered.
\end{itemize}

Through numerical simulations of our model, we analyzed the fractal structure of the aggregates and the cluster size distributions generated.  The fractal dimension $D$ provides information about the static or geometrical properties of a single cluster. It was determined using two different methods, namely, sand box and pair correlation function. The sand box method consists in determining the number of particles $M(r)$ in the cluster inside a hyper-sphere of radius $r$ centered in a given reference point. In general, the following relation is valid

\begin{equation}
\langle M(r) \rangle \sim r^D
\label{eq_dimension}
\end{equation}
where $D$ is the fractal dimension of the cluster. The $\langle \cdot \rangle$ represents an average done over several distinct reference points.

In turn, the pair correlation function method consists in determining the probability to find two points in a cluster separated by $r$. This probability is given by the correlation function

\begin{equation}
c(r) = \frac{1}{S}\sum_{r'}\rho(\vec{r}+\vec{r'})\rho(\vec{r'}),
\label{correlation_function}
\end{equation}
where $S$ is the number of particles comprising the cluster and the sum extends over all cluster particle positions $\vec{r'}$. Here, $\rho$ is the local density, i.e., $\rho(\vec{r})=1$ if at the point $\vec{r}$ there is a particle belonging to the cluster. Typically, the correlation function scales with $r$ as

\begin{equation}
c(r) \sim r^{-\alpha}.
\end{equation}
In addition, the exponent $\alpha$ obeys the relation

\begin{equation}
D = d - \alpha
\end{equation}
in which $d$ is the dimension of the space where the cluster is embedded. In order to minimize boundary effects, the particles taken as references to the measurements performed using both methods are always located within a square region of size $L/2$ centered in the aggregate's mass center.

The dynamical cluster size distribution function $n_s(t)$ is defined as the number of clusters $N_s(t)$ in a unit volume consisting of $s$ particles at time $t$. Then, $n_s(t) = N_s(t)/L^d$ in which $L$ is the linear size of the lattice and $d=2$ is the spatial dimension. The cluster size distribution function conveys information about the evolution of the ensemble of aggregates present in the system. For the original diffusion-limited CCA model (DLCCA), the cluster size distribution function is described by the dynamical scaling relation \cite{Vicsek}

\begin{equation}
n_s(t) \sim s^{-2} f(s/t^z)
\end{equation}
where the functional form of $f(x)$ depends on the exponent $\gamma$ controlling cluster mobility. Specifically, there is a critical value $\gamma_c$ separating two aggregation regimes. For $\gamma < \gamma_c$,

\begin{equation}
f(x) \sim x^2 g(x)
\end{equation}
with $g(x)$ exponentially small for both $x<<1$ and $x>>1$. Thus, $f(x)$ is a bell-shaped curve. In turn, for $\gamma > \gamma_c$, the function $f$ is a monotone curve

\begin{equation}
f(x) \sim \left\{ \begin{array}{ll}
                    x^\delta & \mbox{, if } x<<1,\\
                    <<1 & \mbox{, if } x>>1.
                 \end{array}
          \right.
\end{equation}

\section{Results and discussion}
\label{resultado}
In all simulations the aggregation process starts with $N_0=0.05 \times L^2$ particles and evolves up to the time in which only one cluster remains on the lattice. We have used square lattices of size up to $L=800$ and averages involving up to $100$ independent samples. Two diffusion-limited regimes of aggregation were considered: the size-independent ($\gamma=0$) and the hydrodynamical ($\gamma=-0.5$, $d=2$) cluster migration. Both, the fractal structure of the aggregates and the dynamical scaling of the cluster size distribution, were determined.

\subsection{Model I - The effects of particle replication}

\begin{figure}[hb]
\begin{picture}(0,0)
\put(55,-3){t=20} \put(180,-3){t=640} \put(320,-3){t=2560}
\put(55,-165){t=20} \put(180,-165){t=160} \put(320,-165){t=320}
\end{picture}
~\\ 
\subfigure[]{
\resizebox{4.5cm}{!}{\includegraphics*{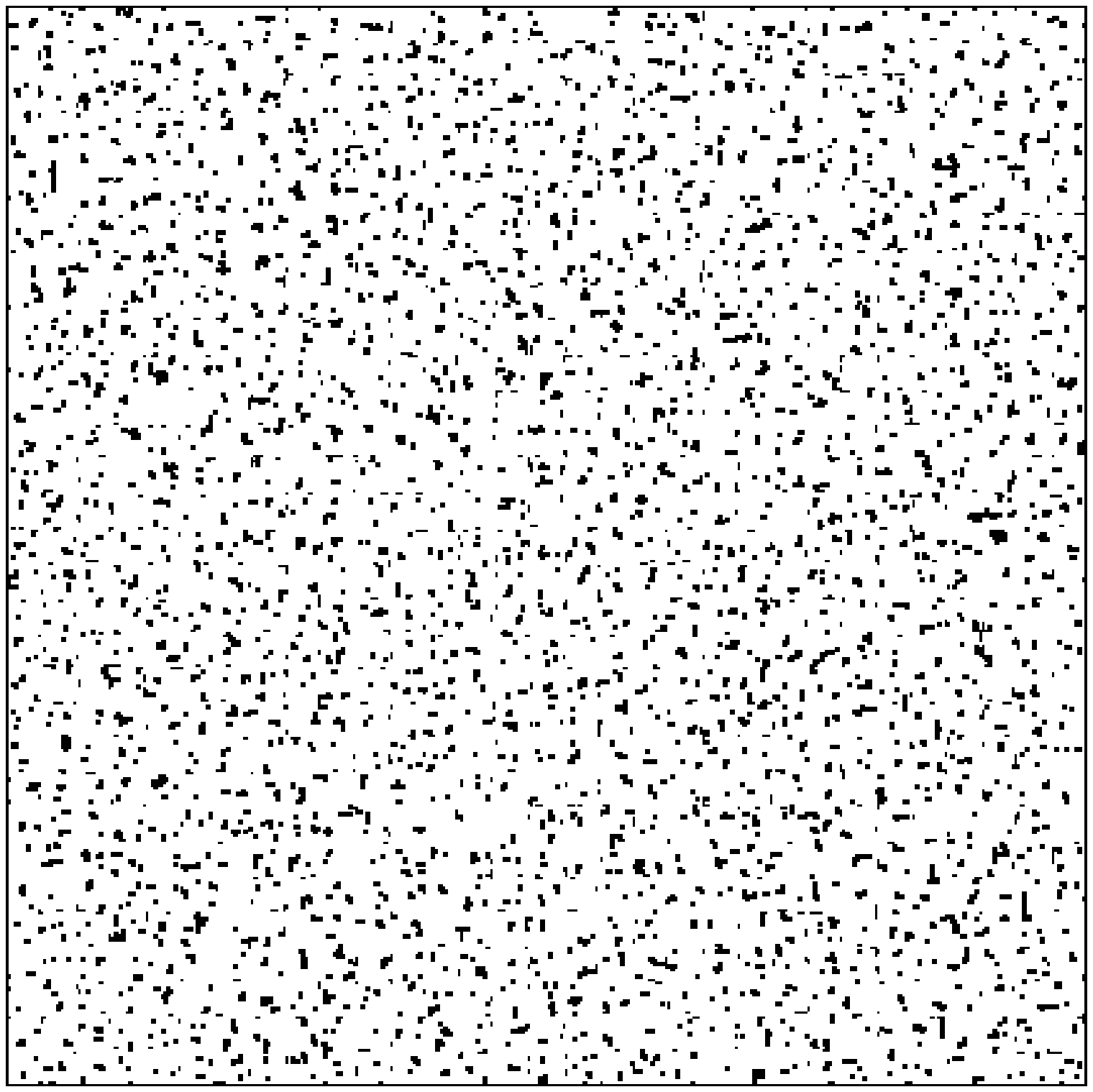}}
\resizebox{4.5cm}{!}{\includegraphics*{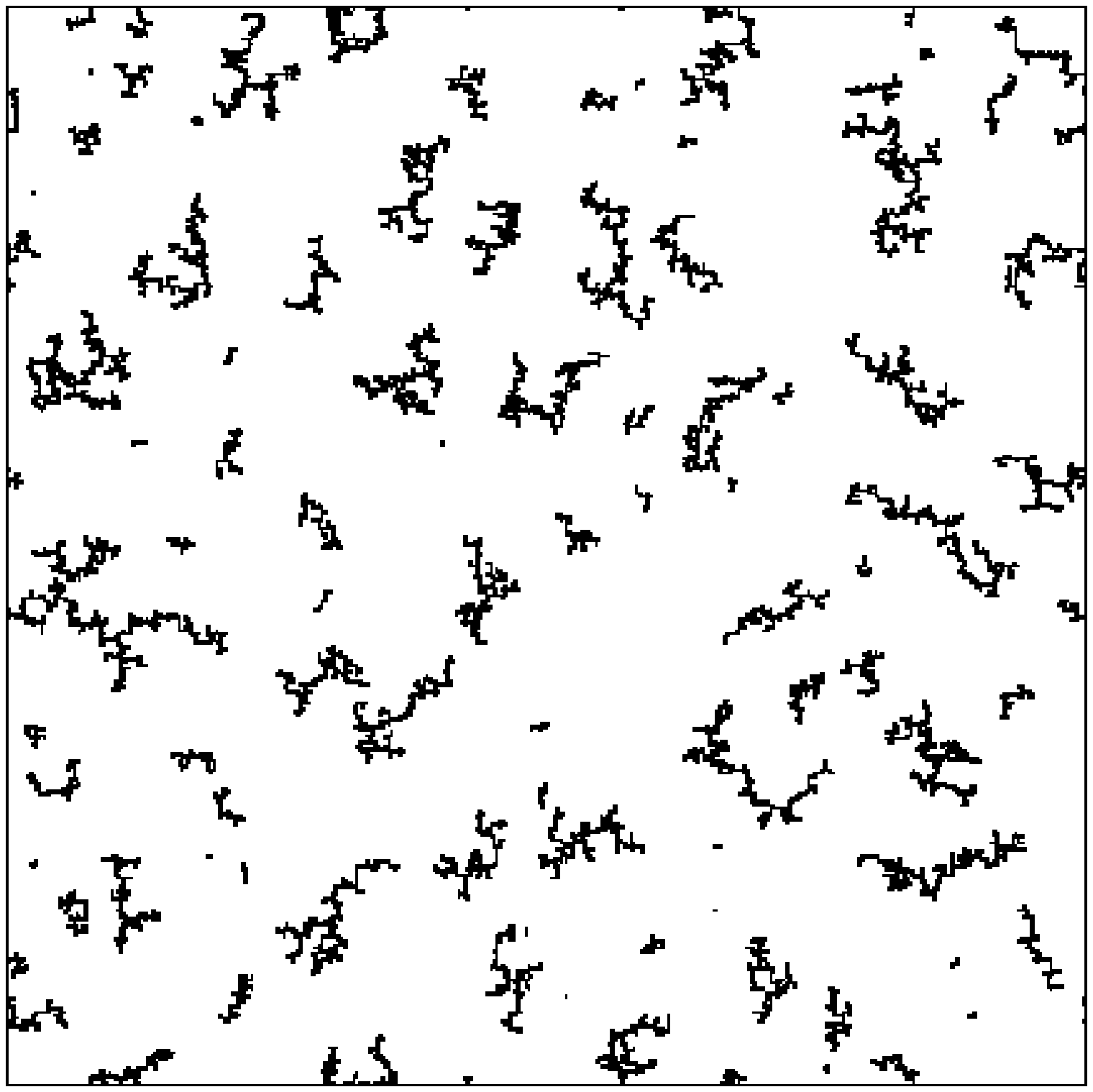}}
\resizebox{4.5cm}{!}{\includegraphics*{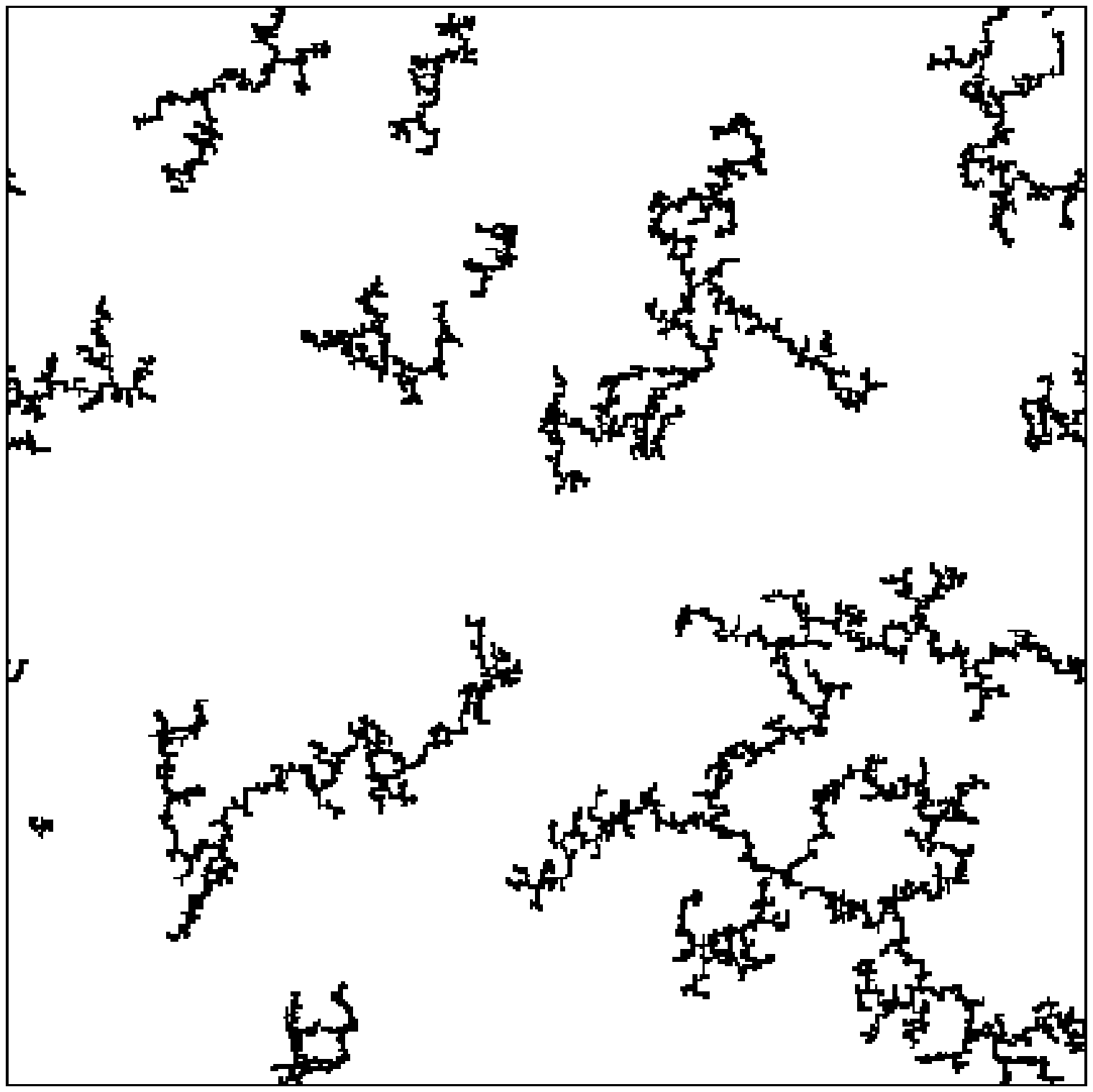}}}\\~\\

\subfigure[]{
\resizebox{4.5cm}{!}{\includegraphics*{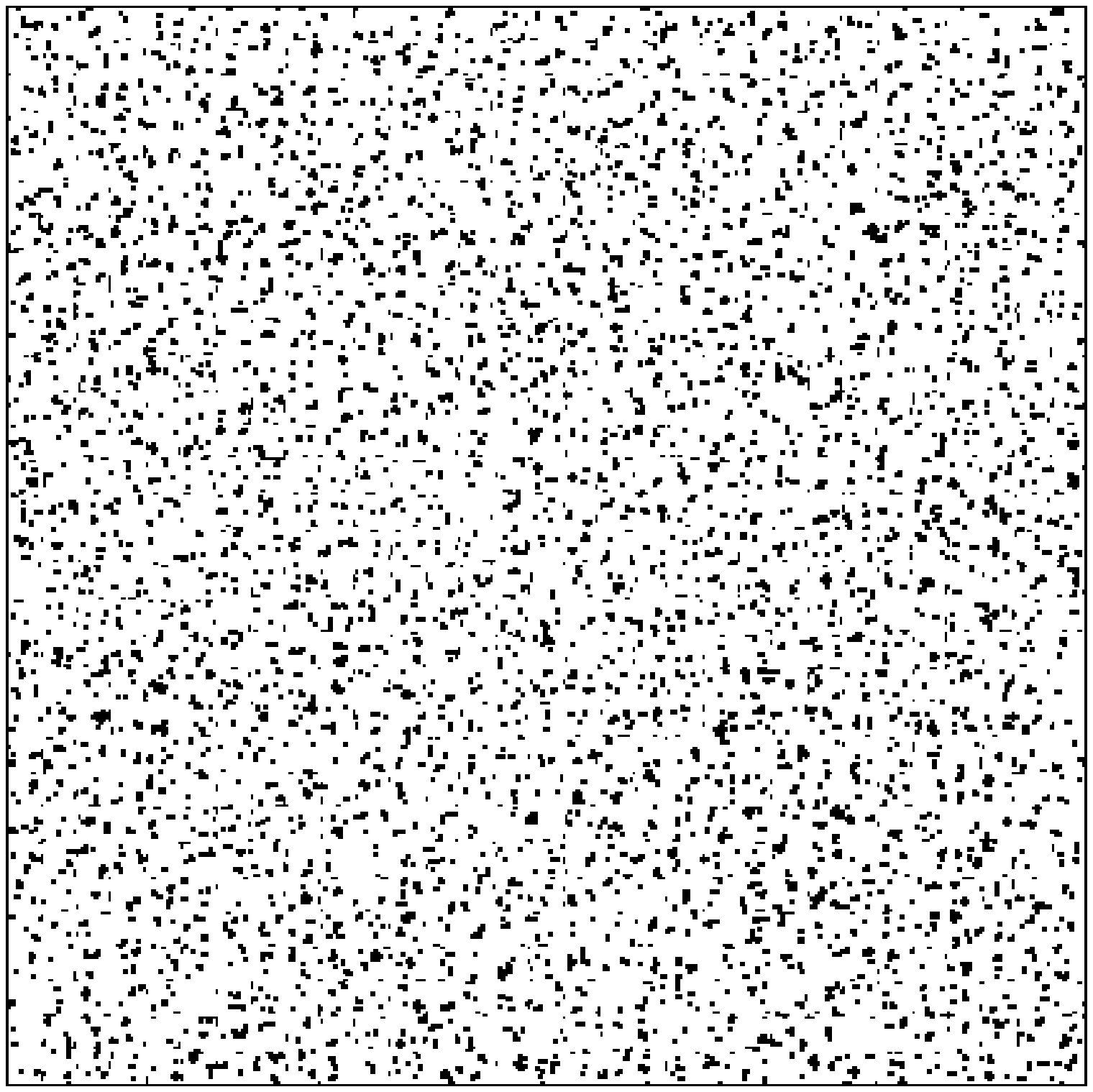}}
\resizebox{4.5cm}{!}{\includegraphics*{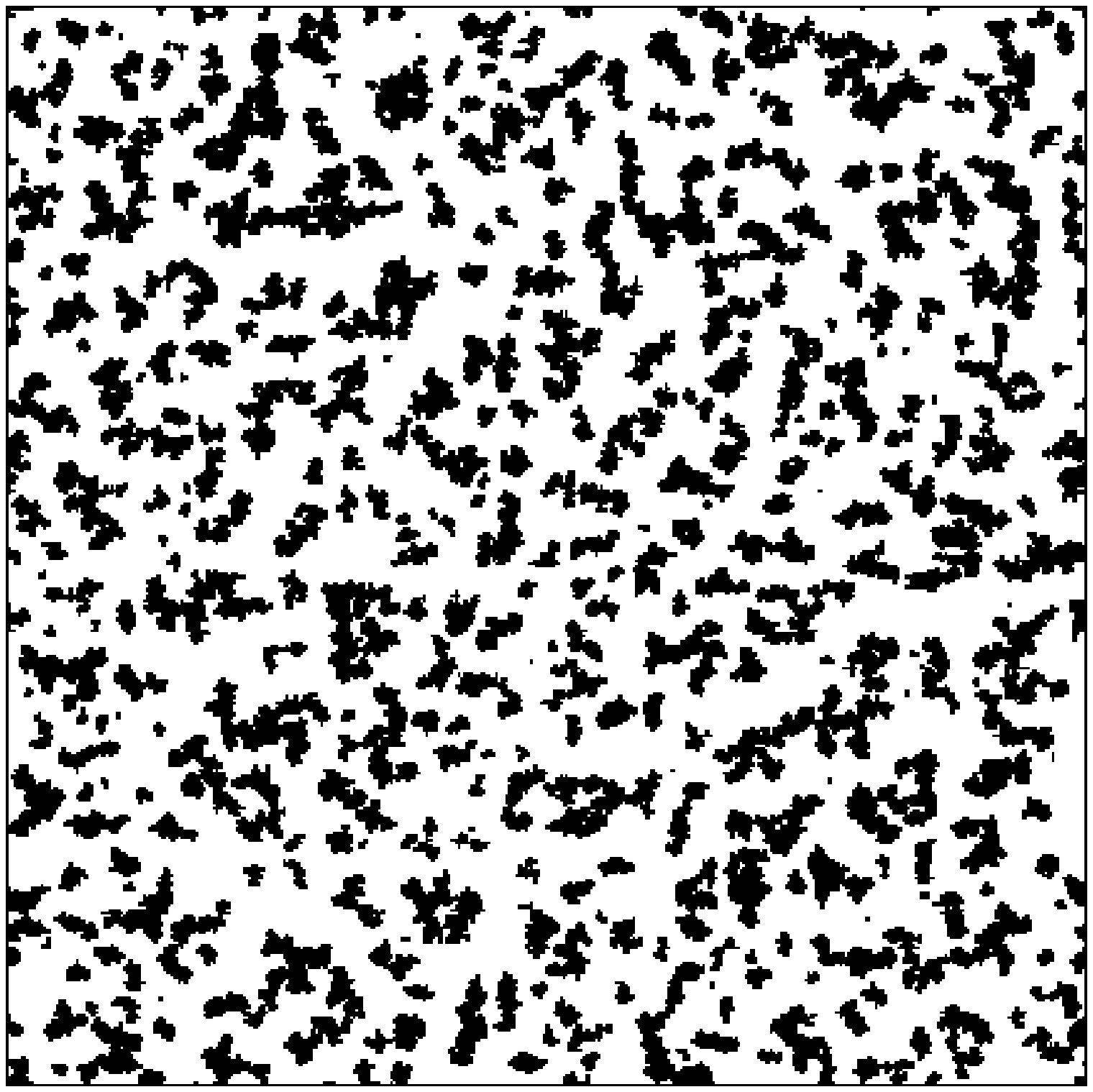}}
\resizebox{4.5cm}{!}{\includegraphics*{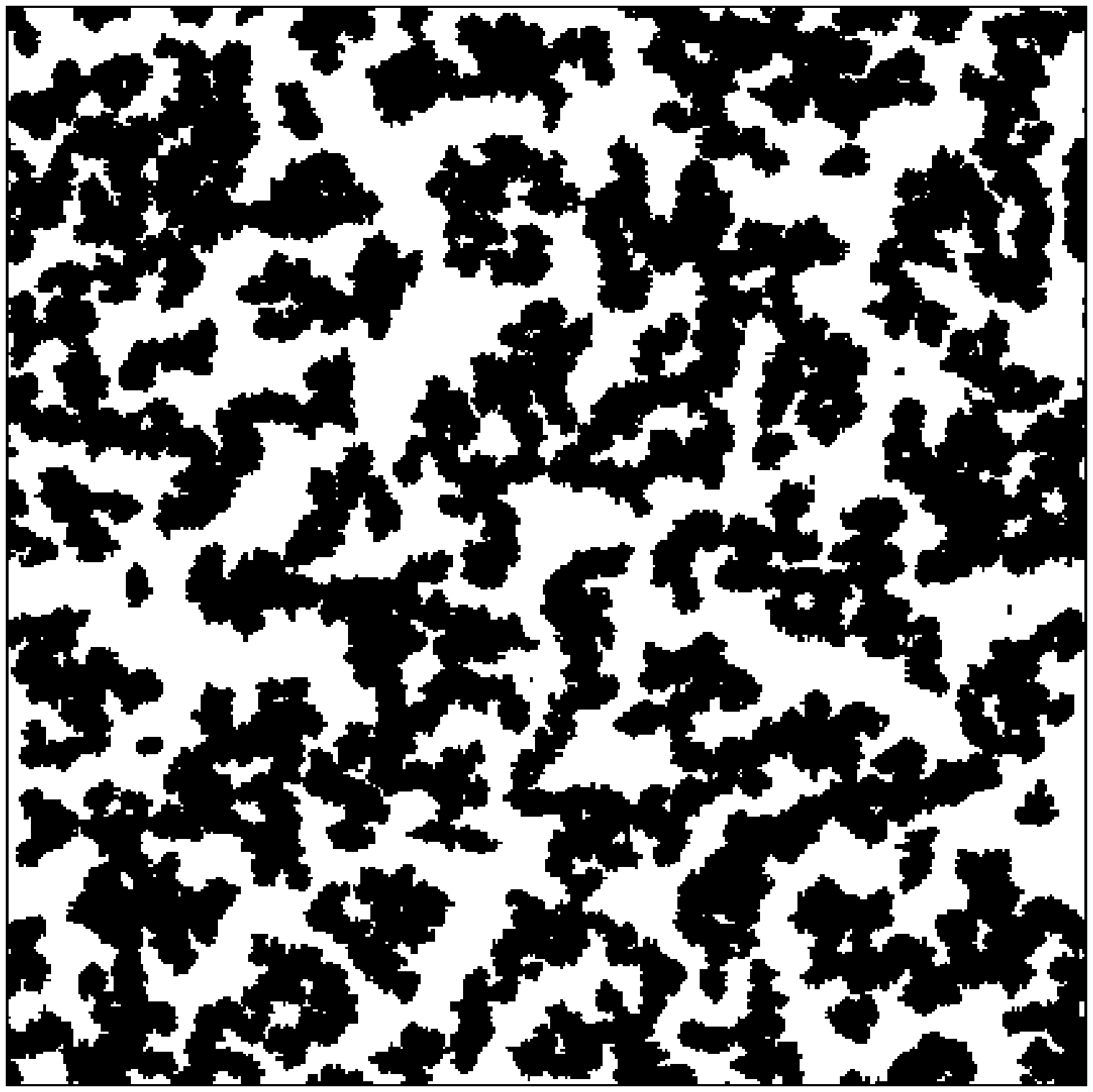}}} \\
\caption{\label{pattern} Aggregation patterns for (a) $p_{div}=0$ (CCA case) and (b) $p_{div}=0.025$, respectively. A square lattice of linear size $L=400$ was used.}
\end{figure}

\begin{figure}[]
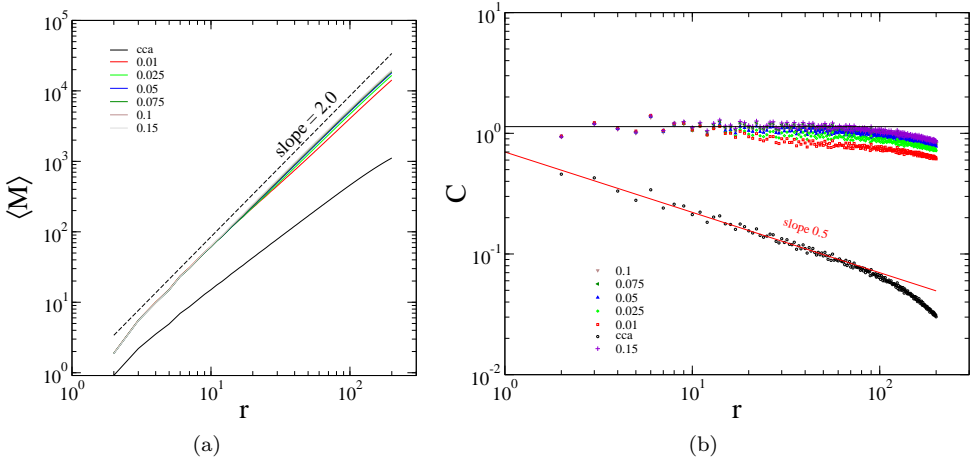

\subfigure[]{\resizebox{5.5cm}{!}{\includegraphics*{d.eps}}}~~
\subfigure[]{\resizebox{7cm}{!}{\includegraphics*{cr.eps}}}
\caption{\label{dimension} (a) The cluster mass $M$ and (b) the pair correlation function $C$ as functions of the radius $r$ for various values of the particle replication rate $p_{div}$.}
\end{figure}

In figure \ref{pattern} are shown snapshots of the aggregation processes for $p_{div}=0$ (the original CCA model) and for a non-null value of $p_{div}$ at distinct times. As expected for cluster-cluster aggregation models, the number of clusters decreases and larger ramified aggregates emerge with increasing time. Furthermore, for $p_{div} \ne 0$ the clusters become fat rapidly as time evolves. This is neatly shown in figure~\ref{dimension}. The fractal dimensions $D$ and exponents $\alpha$ converge to $D=2$ and $\alpha=0$ at long times for various values of $p_{div}$. 
These convergences are very fast, indicating a homogeneous distribution of the mass within larger clusters. However, for very small values of $p_{div}$, a crossover was observed in the cluster mass scaling. For small $r$, the mass scales with an exponent $D=2$, but at larger $r$ this exponent decreases towards $D \simeq 1.45$ for $\gamma=0$ and $p_{div}=0$ \cite{Meakin2}, the original DLCCA value. The asymptotic dimension $D$ observed in this scaling increases with the particle replication rate. Thus, for instance, the fractal dimension of the largest simulated cluster is $D \simeq 1.73$ for $P_{div} = 0.01$.

\begin{figure}
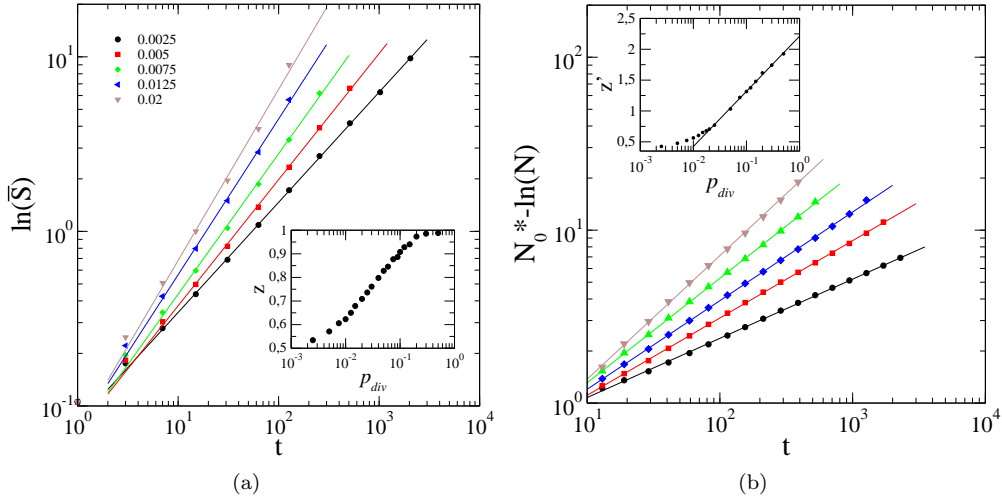

\subfigure[]{\resizebox{6.5cm}{!}{\includegraphics*{s_med.eps}}\label{s_med}}~~
\subfigure[]{\resizebox{6.5cm}{!}{\includegraphics*{Nmed.eps}}\label{n_med}}
\caption{\label{mean_cluster_size} (a) Mean cluster size for five  division rates $p_{div}$ and (b) the average number of clusters in the system as functions of time. The insets show the exponents $z$ and $z'$ as functions of $p_{div}$. In (b) the values for the particle replication rates are the same used in (a).}
\end{figure}

In figure \ref{mean_cluster_size}, the mean cluster size is shown. Differently from the CCA model, where a power law is observed, we found that the mean cluster size increases in time according a stretched exponential for small particle replication rates:

\begin{equation}
\overline{S} = S_0 \exp\left(t^z/ \tau \right),
\label{S_med}
\end{equation}
with the $z$ exponent tending towards $1$, as can be seen in the inset. The characteristic time $\tau$ scales as a power law for large $p_{div}$ ($\tau \sim p_{div}^{-0.75}$). Also, the total number of clusters decreases following a stretched exponential behavior
\begin{equation}
N = N_0 \exp\left(- t^{z'}/\tau' \right).
\end{equation}
Differently from the case of the mean cluster size (inset of the figure \ref{n_med}), we observed that the exponent $z'$ increases logarithmically with the particle replication rate. Again, the characteristic time $\tau'$ scales as a power law of $p_{div}$ ($\tau' \sim p_{div}^{0.93}$).

\begin{figure}
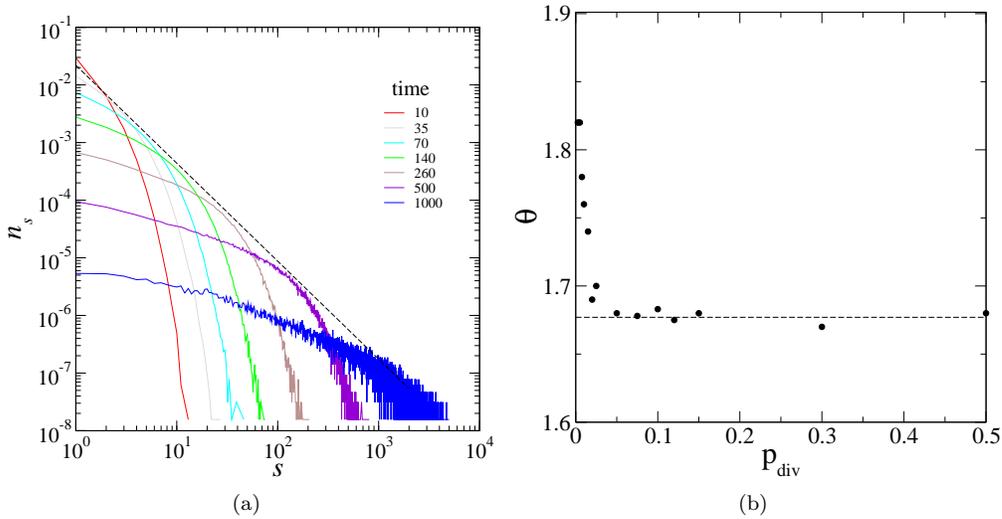

\begin{center}
\subfigure[]{\resizebox{6.5cm}{!}{\includegraphics*{ns_s.eps}}\label{ns_s}}~~
\subfigure[]{\resizebox{6.5cm}{!}{\includegraphics*{theta.eps}}\label{theta}}
\caption{(a) Typical result for the size distribution functions at different times steps obtained using $p_{div}=0.01$ and system size $L=400$. The dashed line is a power law -- the envelope curve. (b) The exponent $\theta$ for different particle replication rates.}
\end{center}
\end{figure}

The dynamical cluster size distribution $n_s(t)$ is shown in figure~\ref{ns_s} at fixed times and small values of the particle replication rate $p_{div}$. At long times, a power law behavior is observed up to a cutoff value of the cluster size $s$. This behavior is analogous to that observed for the original CCA model~\cite{Vicsek}. The distribution functions for different values of the replication rate approach an envelope curve (the dashed line) described by a power law decay,

\begin{equation}
n^*_s \sim {s^*}^{-\theta}.
\end{equation}
Here, $n^*_s$ is the value of the distribution function and $s^*$ is the cluster size at the cut off of the curves in the figure \ref{ns_s}. As shown in the figure \ref{theta}, the exponent $\theta$ approaches a constant value $\overline\theta = 1.677(3)$. We can see from figures \ref{s_med} and \ref{theta} that the $z$ exponent converges to a value very close to $1$ (representing an exponential growth of the mean cluster size) at the same division rate in which $\theta$ reaches its constant value. This indicates that the dynamical aggregation process is dominated mainly by the replication of the particles beyond a characteristic $p_{div}$ value.

\begin{figure}
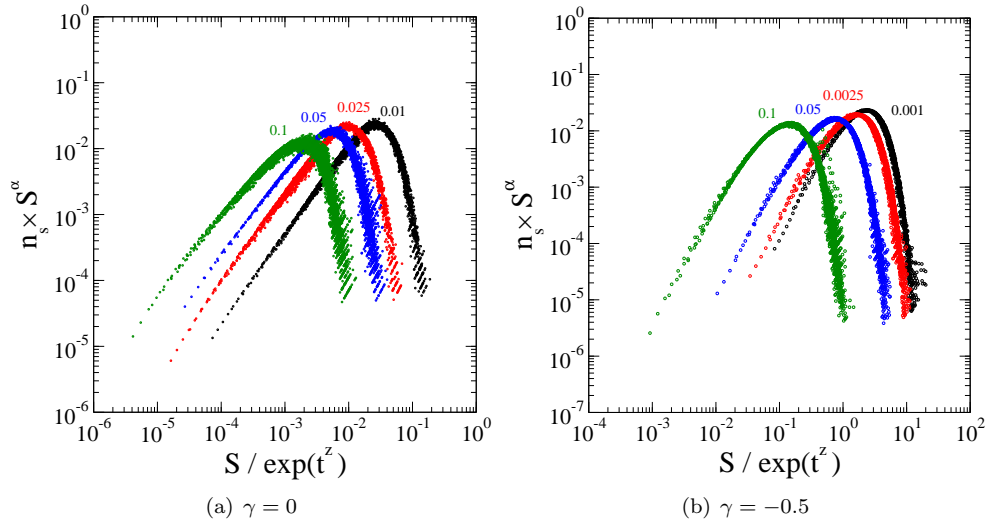

\begin{center}
\subfigure[$\gamma=0$]{\resizebox{6.45cm}{!}{\includegraphics*{cols_gamma0.eps}}}
\subfigure[$\gamma=-0.5$]{\resizebox{6.45cm}{!}{\includegraphics*{cols_gamma05.eps}}}
\caption{\label{colapso}The collapse of the size distribution functions obtained using the ansatz in equation \ref{scale} for (a) $\gamma = 0$ (the mass-independent cluster motility) and (b) $\gamma = -0.5$. Several values of the division rate were used.}
\end{center}
\end{figure}

As observed for the original CCA model, the cluster size distribution for the aggregation with particle replication can be well represented by a scaling function. In this case, we have found the following ansatz

\begin{equation}
n_s(t) \sim s^{-\alpha} f\left(\frac{s}{\exp(t^z)}\right) 
\label{scale}
\end{equation}
where the exponent $\alpha$ is determined from the power law envelope in the figure \ref{ns_s}. The scaling function $f(x)$ follows a power law for  $x \ll 1$ and decreases rapidly for $x \gg 1$. As shown in figure \ref{colapso}, a very good collapse was obtained using the scaling assumption prescribed in equation \ref{scale}. It is important to mention that, as can be seen in the figure \ref{colapso}(b), the same scaling function was obtained for the case where $\gamma = -0.5$.

In summary, the particle replication changes not only the fractal dimension of the aggregates, as occurs when cluster rotation \cite{Meakin3}, internal restructuring \cite{Connor} or other mechanisms are introduced in the CCA model, but also the dynamics of the aggregation kinetics. Indeed, the mean cluster size, the total number of clusters and the cluster size distribution functions exhibit stretched exponential dependences in time. Furthermore, the parameters (exponents and characteristic time scales) controlling such stretched exponentials are related to $p_{div}$, the particle replication rate. However, the scaling of $n_s(t)$ does not explain the experimental data observed for normal and tumoral cell lineages grown in culture. In fact, in the range of small cluster sizes observed in monolayer cultures, the simplest cell aggregation regimes exhibit $n_s(t)$ decaying as power laws with decreasing exponents (slower decays) as the time in culture increases \cite{Rose,Vilela}. This is the case for HN-5, MCF-10, and MB-231 cells, for instance. In contrast, as we have found, particle replication does not introduce a time dependence in the exponent $\alpha$ describing the decay of $n_s$ at a fixed time. So, it seems that additional mechanisms must be taken into account in order to describe adequately even this simplest cell aggregation kinetics in culture. Cell signaling appears as the major of such biological mechanisms, since gradients of cell-cell signals give rise to chemotactically driven cluster motility. Chemotaxy alters the nature of cluster diffusion in the CCA model.

\subsection{Models II and III - The effects of Chemotaxis}

\begin{figure}[b]
\subfigure[Original CCA]{\resizebox{5cm}{!}{\includegraphics*{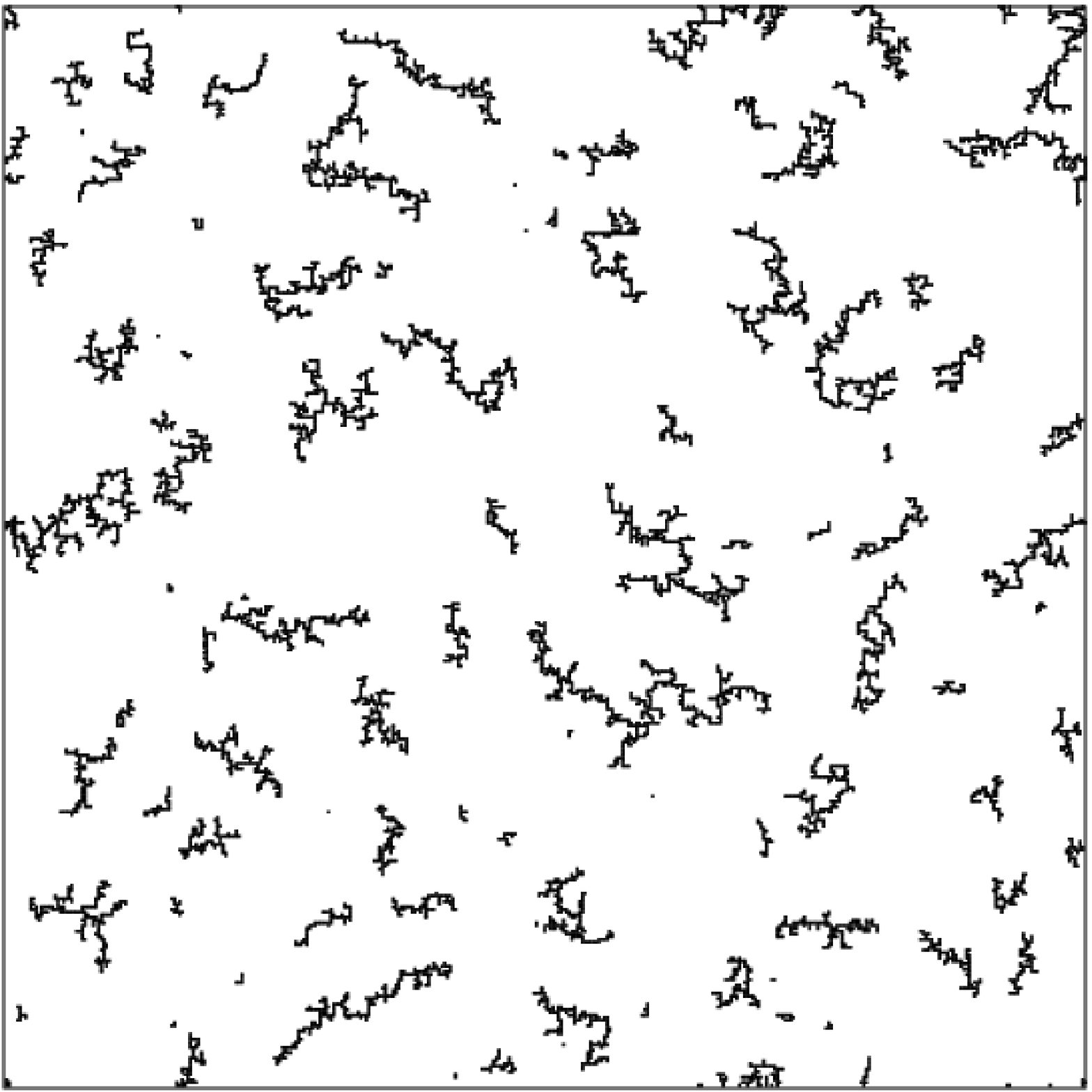}}} ~~~~
\subfigure[Model I]{\resizebox{5cm}{!}{\includegraphics*{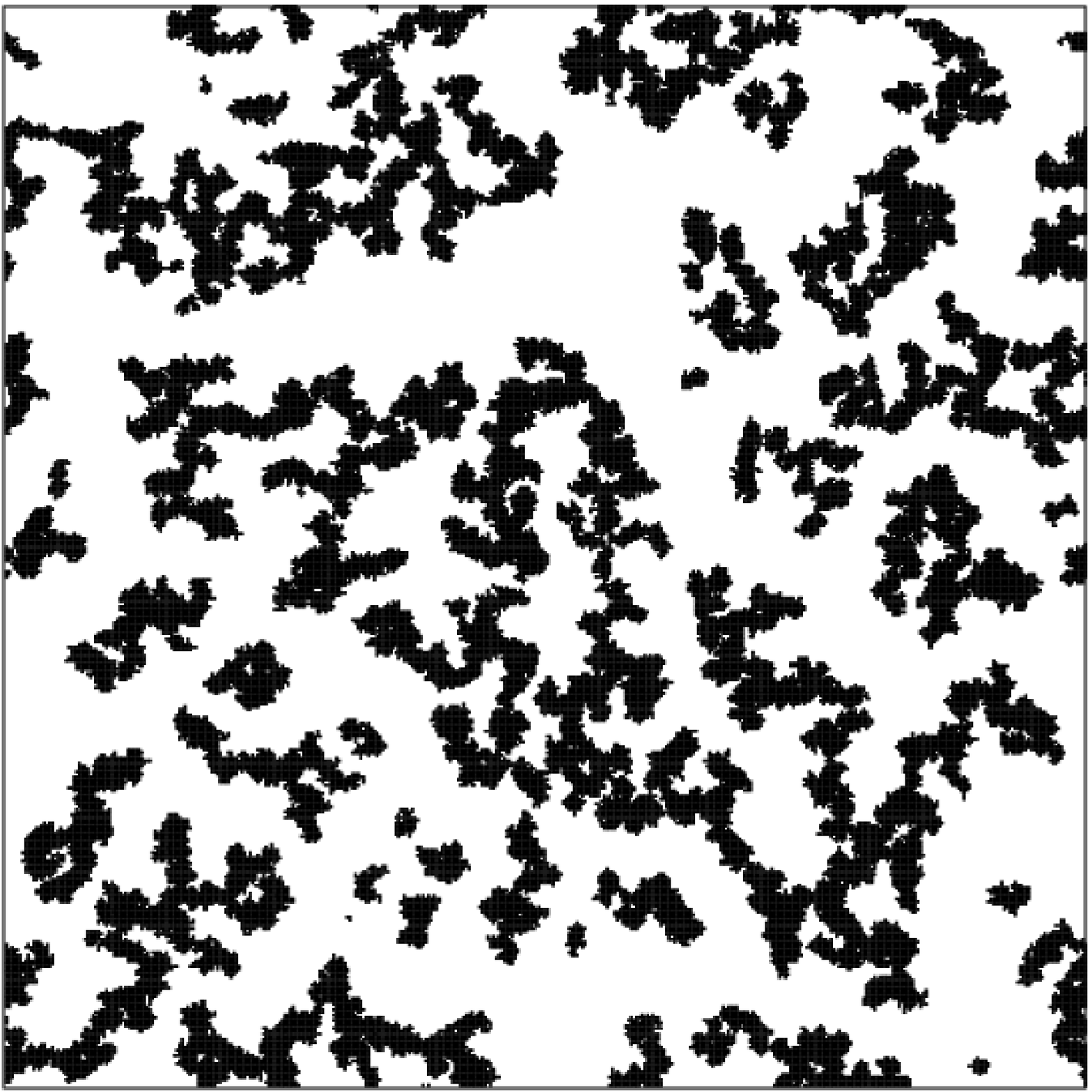}}}\\
\subfigure[Model II]{\resizebox{5cm}{!}{\includegraphics*{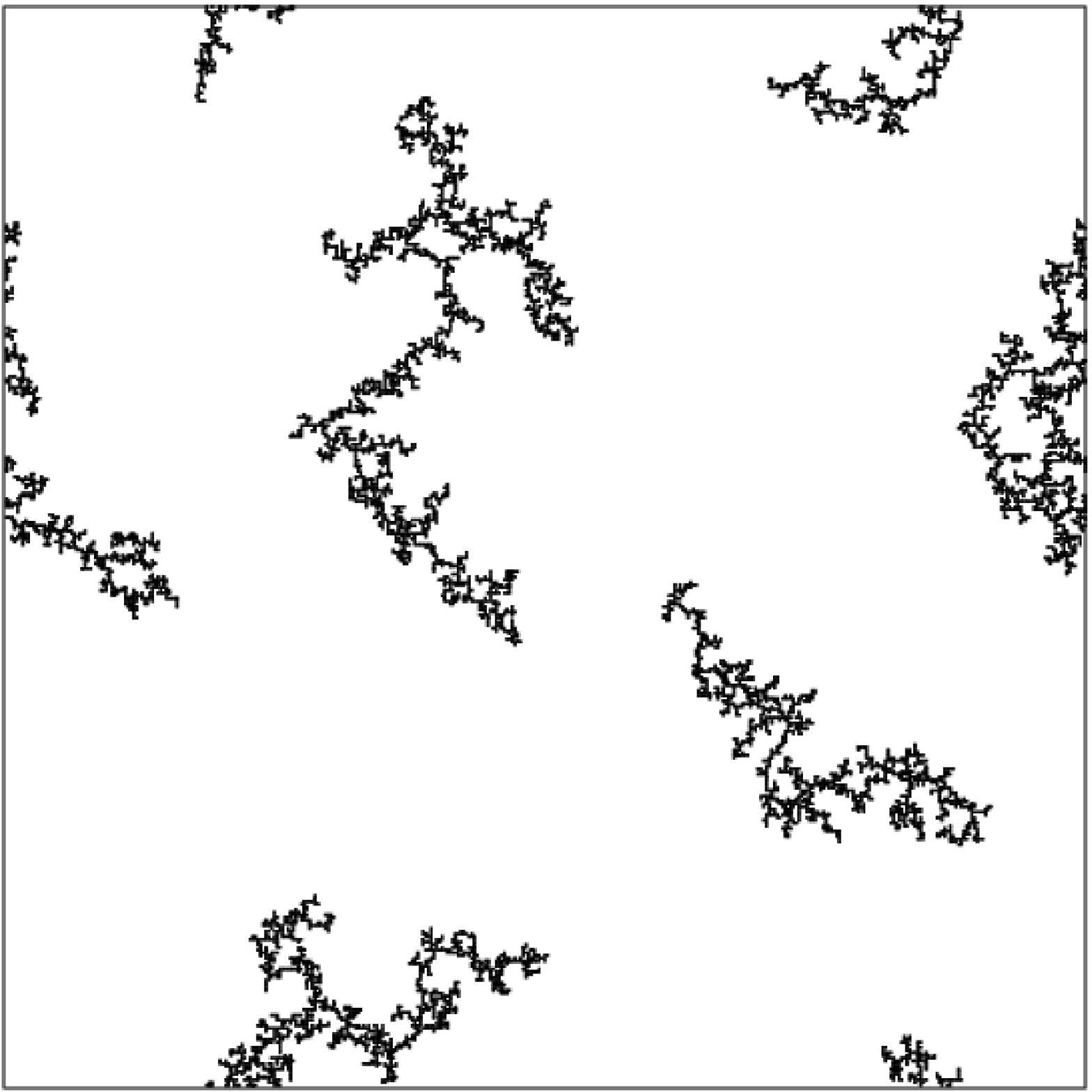}}} ~~~~
\subfigure[Model III]{\resizebox{5cm}{!}{\includegraphics*{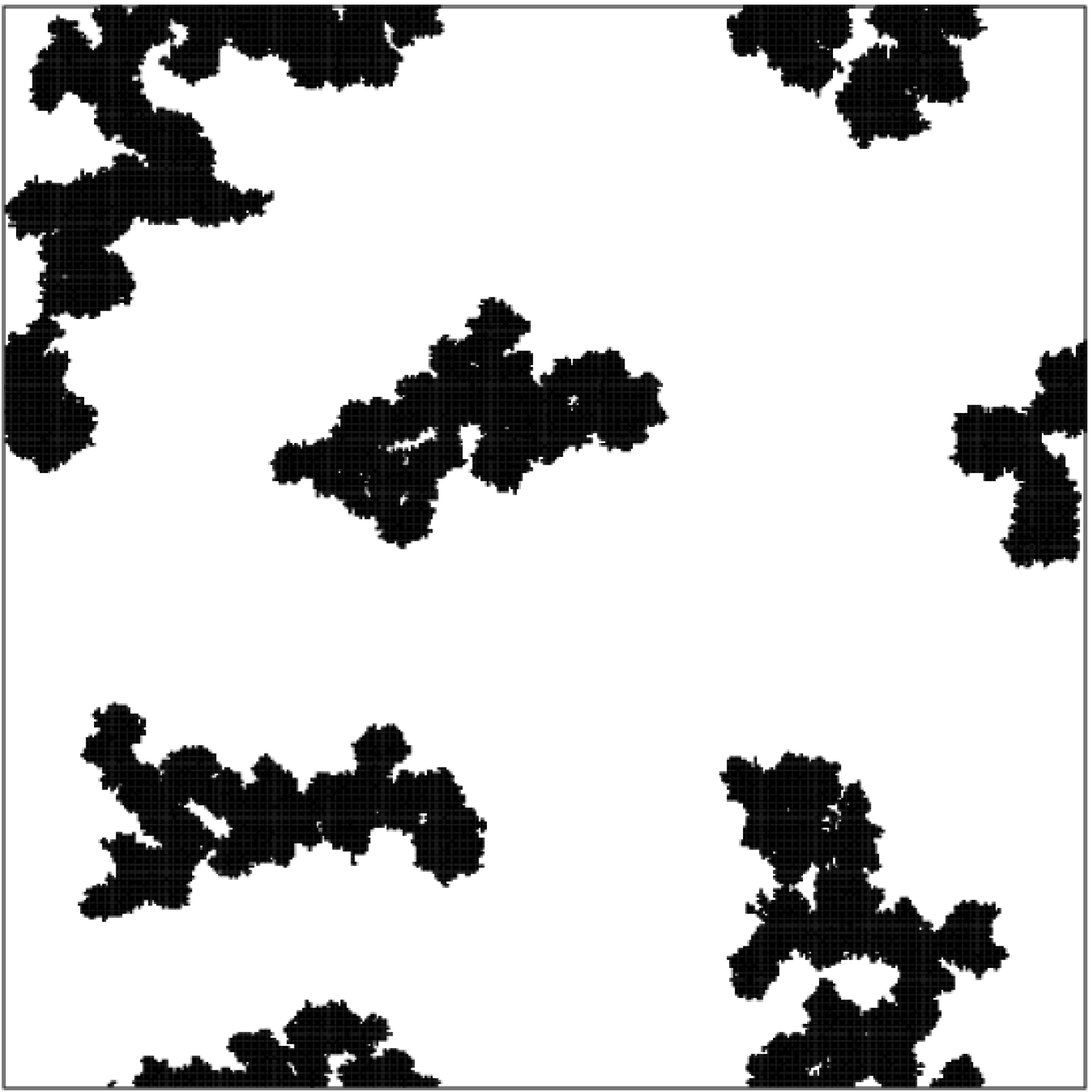}}} \\
\caption{\label{pattern1} Aggregation patterns after $1200$ time steps for (a) original CCA model, (b) Model I with the rate replication $p_{div} = 0.01$, (c) Model II and (d) Model III with the rate replication $p_{div} = 0.01$. A square lattice of linear size $L=400$ was used.}
\end{figure}

Now we consider the effects of cell signaling on the kinetic aggregation of cells in culture. Chemotactically driven motility is taken into account in the model versions II and III. In the figure \ref{pattern1} snapshots of the aggregation patterns generated by different model versions after $1200$ time steps are shown. As one can easily see, chemotaxy drastically changes the patterns formated. A comparison between figures \ref{pattern1}(a) and (c) neatly reveal that aggregation rate is significantly enlarged under the action of a diffusive chemoattractant. Furthermore, for a fixed time the sizes of the aggregates grown under chemotaxy are larger than those 
generated without cell-cell signaling. However, the morphology of these chemotactically driven aggregates is the same observed for the original CCA clusters. Indeed, the fractal dimension of the clusters emerging from the model II equals that of the CCA model.

\begin{figure}[hbt]
\resizebox{6.5cm}{!}{\includegraphics*{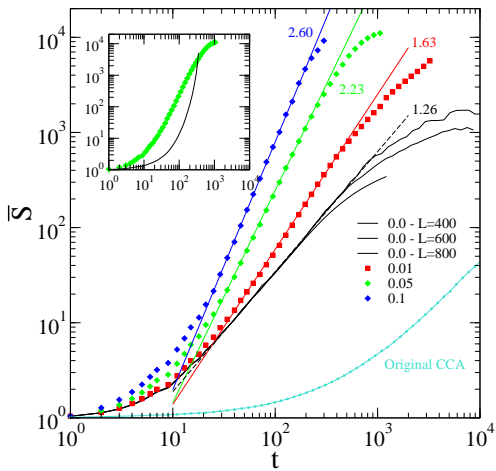}}

\caption{\label{smed_chemotaxix} Mean cluster size for the model versions II and III. Three different system size are use for the model II to show the finite size effects. In the case of the version III we used three values of division rate. The inset shows the curves for $p_{div}=0.05$ using the version II and III.}
\end{figure}

Again, a comparison between figures \ref{pattern1}(b) and (d) also reveal that both the aggregation rate and the cluster sizes are greatly enhanced primarily due to chemotaxy. Moreover, particle replication. neatly affects the cluster morphology leading to fat aggregates with fractal dimensions $D \rightarrow \infty$. The evolution in time of the mean cluster size $\bar{S}$ for all these model versions is shown in figure \ref{smed_chemotaxix}. It is worthy to notice that, after an initial transient, $\bar{S}$ increases slowly in a chemotactic driven regime than without intercellular signaling, as shown in the inset of figure \ref{smed_chemotaxix}. Indeed, the number of peripheral particles, those able to replicate, is much larger in the absence of a chemoattractant whose effect is promote a fast merging of separated clusters.

\begin{figure}[hbt]
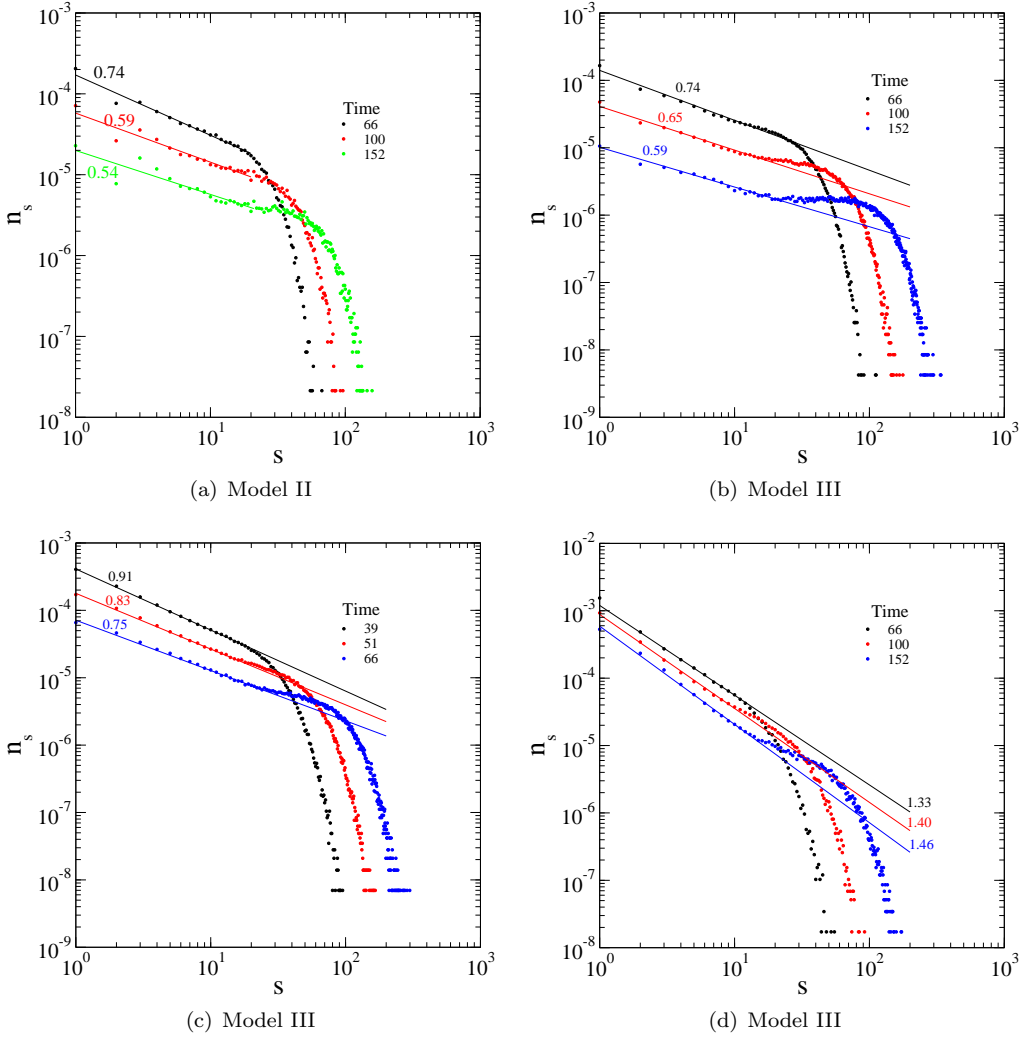

\subfigure[Model II]{\resizebox{6.5cm}{!}{\includegraphics*{ns_s_pdiv0g0.eps}}} ~~~~
\subfigure[Model III]{\resizebox{6.5cm}{!}{\includegraphics*{ns_s_pdiv001g0.eps}}} \\
\subfigure[Model III]{\resizebox{6.5cm}{!}{\includegraphics*{ns_s_pdiv005g0.eps}}} ~~~~
\subfigure[Model III]{\resizebox{6.5cm}{!}{\includegraphics*{ns_s_pdiv001g05.eps}}}\\

\caption{\label{results_chemotaxix} Cluster size distribution functions for the model versions II and III for different times. Two distinct particle replication rates ((b) $p_{div}=0.01$  and (c) $0.05$) and cluster size dependent motility ((d) $\gamma = - 0.5$ for $p_{div}=0.01$) were used for model III.}
\end{figure}

Finally, the cluster size distribution functions correspondent to the models II and III are shown in figure \ref{results_chemotaxix}. The major result is that chemotactically driven motility leads to distributions $n_s(t)$ decaying as power laws with exponents that decrease as the time increases in the case of cluster size independent motility ($\gamma=0$). Just the same qualitative behavior exhibited by several cell lineages in culture. Cell division does not changes this power law scaling as shown in figures \ref{results_chemotaxix}(b) and (c). It is possible that particle replication introduces corrections to these power laws hardly detectable at the scale of our present simulations. Nevertheless, particle replication is an essential feature to generate clusters with shapes similar to those observed in cell aggregates growing in monolayer culture. In contrast, figure \ref{results_chemotaxix}(d) shows that for cluster size dependent motility ($\gamma=-0.5$) the exponents controlling the power law scaling of $n_s(t)$ increases with time. This occurs because the larger clusters become progressively less mobile impairing the formation of even larger aggregates by the merging of such clusters. As a consequence, the mean cluster size increases more slowly than in the case of size independent motility.

Despite its partial success, we must emphasize that the present model is a first order approach to the complex kinetics of cell aggregation in culture. This model can not explain neither cluster size distributions with characteristic sizes nor the transitions between kinetic regimes controlled by power-law and exponentially distributed $n_s(t)$, as observed in culture assays. At least another main feature, cell-cell adhesion, seems to be required. Adhesion events can promote cluster restructuring through the formation of new cell contacts, modulate collective cell motility and control cluster fragmentation. A sticking (adhesion) probability which changes dynamically in response to alterations in the growth conditions seems to be involved in the phenotypic transitions observed in cell aggregation kinetics \cite{Vilela}. Currently, we are extending the present model to take into account chemotaxy and cell-cell adhesion that driven cluster formation and collective cell motility.

\section{Conclusions}
\label{conclusao}
A cluster-cluster aggregation model with stochastic particle replication was investigated as a model for the growth of animal cells in monolayer cultures. Specifically, the dynamics of cell migration, aggregation and division was modeled through a CCA model with particle replication. The scaling laws generated by this extended CCA process were determined and compared with those observed in culture assays. The mean cluster size, the total number of clusters and the cluster size distribution functions exhibit stretched exponential dependences in time with exponents and characteristic times fixed by the particle replication rate. However, particle replication alone cannot account for the cluster size distributions of cellular aggregates grown in culture. Neatly, complex biological phenomena such as chemotaxis, haptotaxis, cell-cell adhesion, collective cell polarization and migration, that were neglected in the present model, must be included in a reliable model for cell aggregation in culture.

\ack This work was partially supported by the Brazilian Agencies CAPES, CNPq, and FAPEMIG. We thank to L. R. Paiva for the critical reading of the manuscript.


\section*{References}
\begin{thebibliography}{99}

\bibitem{Rose} Mendes R L, Santos A A, Martins M L and Vilela M J 2001 \textit{Cluster size distribution of cell aggregates in culture} Physica A \textbf{298} 471-487

\bibitem{Vilela} Vilela M J, Martins M L, Renato N S, Cazares L, Lattanzio F, Ward M and Semmes O J 2007 \textit{Proteomic and fractal analysis of a phenotypic transition in the growth of human breast cells in culture} J. Stat. Mech. P12006 1-18

\bibitem{Rubin} Rubin H 2001 \textit{Selected cell and selective microenvironmental in neoplastic development} Cancer Res. \textbf{61} 799-807

\bibitem{Chow} Chow M and Rubin H 2000 \textit{Clonal selection versus genetic instability as the driving force in neoplastic transformation} Cancer Res. \textbf{60} 6510-6518

\bibitem{Chambers} Chambers A F, Groom A C and MacDonald I C 2002 \textit{Dissemination and growth of cancer cells in metastatic sites} Nature Rev. Cancer \textbf{2} 563-572

\bibitem{Keller} Keller R 2005 \textit{Cell migration during gastrulation} Curr. Opin. Cell Biol. \textbf{17} 533-541

\bibitem{Friedl} Friedl P and Br\"ocker E B 2000 \textit{The biology of cell locomotion within three-dimensional extracellular matrix} Cell Mol. Life Sci. \textbf{57} 41-64

\bibitem{Enmon1} Enmon Jr. R M, O'Connor K C, Song H, Lacks D J, Schwartz D K, and Dotson R S 2001 \textit{Dynamics of spheroid self-assembly in liquid-overlay culture of DU 145 human prostate cancer cells} Biotechnol. Bioeng. \textbf{72} 579-591

\bibitem{Enmon2} Enmon Jr. R M, O'Connor K C, Song H, Lacks D J, and Schwartz D K 2002 \textit{Aggregation kinetics of well and poor differentiated human prostate cancer cells} Biotechnol. Bioeng. \textbf{80} 580-588

\bibitem{Meakin} Meakin P 1983 \textit{Formation of fractal clusters and networks by irreversible diffusion-limited aggregation} Phys. Rev. Lett. \textbf{51} 1119-1122

\bibitem{Kolb} Kolb M, Botet J and Jullien R 1983 \textit{Scaling of kinetically growing clusters} Phys. Rev. Lett. \textbf{51} 1123-1126

\bibitem{Connor} Song H, Jain K S, Enmon R M, and O'Connor K C 2004 \textit{Restructuring dynamics of DU 145 and LNCaP prostate cancer spheroids} In Vitro Cell. Dev. Biol. --- Animal \textbf{40} 262-267

\bibitem{Harada} Harada T, Isomura A, and Yoshikawa K 2008 \textit{Contraction-induced cluster formation in cardiac cell culture} Phys. D \textbf{237} 2787-2796

\bibitem{Kim} Kim S H J, Yu W, Mostov K, Matthay M A, and Hunt C A 2009 \textit{A computational approach to understand in vitro alveolar morphogenesis} PLoS ONE \textbf{4} e4819

\bibitem{Sandro} Vilela M J, Martins M L and Boschetti S R 1995 \textit{Fractal patterns for cells in culture} J. Pathol. \textbf{177} 103-107

\bibitem{Vicsek} Vicsek T, 1992 \textit{Fractal Growth Phenomena} (Singapore: World Scientific) pp 226-237

\bibitem{Meakin2} Meakin P 1985 \textit{Computer simulation of growth and aggregation processes}, in: On Growth and Form, edited by Stanley H E and Ostrowsky N (Martinus Nijhoff: Dordrecht) pp 111-135

\bibitem{Meakin3} Meakin P and Jullien R 1985 \textit{Structural readjustment effects in cluster-cluster aggregation} J. Physique \textbf{46} 1543-1552

\end{thebibliography}
\end{document}